\documentclass[12pt,a4]{article}
\usepackage{graphicx}
\usepackage{bm}

\title{The Zenith Angle Distribution of Fully Contained Events in 
SuperKamiokande and the Impact of Quasi Elastic Scattering on their Direction}

\author{E. Konishi$^{1}$, Y. Minorikawa$^{2}$, V.I. Galkin$^{3}$,\\
M. Ishiwata$^{4}$ and A. Misaki$^{5,6}$}
\begin{document}
\maketitle
\small{
\noindent{ 
$^1$ Department of Electronics and Information System Engineering, 
    Hirosaki University, 036-8561, Hirosaki, Japan\\
$^2$ Department of Science, School of Science and Engineering, 
    Kinki University, Higashi-Osaka, 577-8502, Japan\\
$^3$ Department of Physics, Moscow State University, 119992, Moscow, 
    Russia\\
$^4$ Department of Physics, Saitama University, 338-8570, Saitama, Japan\\
$^5$ Advanced Research Institute for Science and Engineering, 
    Waseda University, 169-0092, Tokyo, Japan\\
$^6$ Innovative Research Organization for the New Century, 
    Saitama University, 338-8570, Saitama, Japan\\
    }

\begin{abstract}
\small{ 
Quasi Elastic Scattering (QEL) is the dominant mechanism for producing
both {\it Fully Contained Events} and {\it Partially Contained Events} in the
SuperKamiokande (SK) detector for atmospheric neutrinos in the
range $\sim$0.1~GeV to $\sim$10~GeV. In the analysis of SK events, it is assumed
that the zenith angle of the incident neutrino is the same as that of the
detected charged lepton.

In the present paper, we derive the distribution function for the
scattering angle of charged leptons, the average
scattering angles and their standard deviation due to QEL.
Thus, it is shown that the SK assumption for the scattering
angle of the charged lepton in QEL is not valid.
Further, for the rigorous analysis of the experimental data,
the correspondence between the
zenith angle of the charged leptons and that of the incident
neutrinos should be examined by taking account of the influence of the azimuthal angle of
the charged particle on its zenith angle.  We conclude that it is
not possible to reliably assign the zenith angle of the incident
neutrino to that of the charged lepton, and so the zenith angle
distribution of charged leptons under the SK assumption does not reflect that of the real incident neutrinos.

This result has clear implication for attempts to detect neutrino
oscillations from the analyses of {\it Fully Contained Events} and {\it Partially Contained Events} in SuperKamiokande.
}
\end{abstract}

\section{Introduction}

   The report of oscillations between muon and tau neutrinos for
atmospheric neutrinos detected with SuperKamiokande (SK, hereafter) is
claimed to be robustly established for the following reasons:

(1) The discrimination between electrons and muons in the SK energy
range, say, $\sim$0.1 GeV to $\sim$1 GeV, has been proved to be almost perfect, as
demonstrated by calibration using accelerator beams \cite{r1}.

(2) The analysis, for both {\it Fully Contained Events} and 
{\it Partially Contained Events}, of the zenith angle distribution of
electron-like (single ring) events and muon-like (single ring)
events, based on the well established discrimination procedure
mentioned in (1), reveals a significant deficit of muon-like events
but the expected level of electron-like events. It is concluded that muon
neutrinos oscillate into tau neutrinos which cannot be detected due to
the small geometry of SK \cite{r2}.
The analysis of SK data presently yields
${\rm sin}^22\theta > 0.89$ 
and 
$1.8\times 10^{-3}{\rm eV}^2<\Delta m^2 < 4.5\times 10^{-3}{\rm eV}^2$ 
at 90\% confidence level.
\footnote{The numerical values of the combination of ${\rm sin}^22\theta$
and $\delta m^2$ change from time to time. Here, we cite those 
from the most recent work \cite{r3}.}

(3) The analysis of {\it Upward Through Going Events} and 
{\it Stopping Events}, in which the neutrino interactions occur outside
the SK detector, leads to similar results to (2). The charged
leptons which are produced in these categories are regarded as being
exclusively muons, because electrons have negligible probabilities to
produce such events as they lose energy very rapidly in the sorrounding rock.

Thus for these events the discrimination procedure described in (1) is
not required, and ,therefore, the analysis here is independent of the analysis in
(2). For these events, however, the SK group obtains the same parameters for
neutrino oscillations as in (2).
\footnote{It seems strange for different quality
experimental data to give similarly precise results, because {\it
Fully Contained Events} 
are of higher experimental quality compared with those of
both {\it Upward Through Going Muon} and {\it Stopping Muon
Events}. See also footnote 1.}

The most robust evidence for neutrino oscillations is regarded as being
that from the analysis of {\it Fully Contained Events}, because all details of the events are measured inside the detector and its analysis is free of the ambiguities that arise in the other analyses.
 {\it Partially Contained Events} have uncertainties due to
their unknown ending points, and {\it Stopping Muon Events} and {\it
Upward Through-Going Muon Events} both have uncertainties due to the
unknown starting points, preventing a complete analysis of the events
concerned.

In the SK experiment, the direction of the incident neutrino is
assumed to be the same as that of the emitted charged lepton, i.e.,
the (anti-)muon or (anti-)electron, in the analysis of both {\it Fully
Contained Events} and {\it Partially Contained
Events} \cite{r3,r4}. However, as we will show,
this assumption does not hold in the most important energy region,
$\sim$0.1 GeV to $\sim$10~GeV, for
both {\it Fully Contained Events} and {\it Partially Contained
Events}, 
where Quasi Elastic Scattering (QEL)
of the neutrino interaction is most dominant \cite{r5}. In the
present paper, we examine the implication of this assumption in a
quantitative way.

\section{Cross Sections of Quasi Elastic Scattering in the Neutrino 
Reaction and the Scattering Angle of Charged Leptons.}

    We examine the following reactions due to the charged current interaction
(c.c.). \\
   \begin{eqnarray}
         \nu_e + n \longrightarrow p + e^-  \nonumber\\
         \nu_{\mu} + n \longrightarrow p + \mu^- \nonumber\\
         \bar{\nu}_e + p \longrightarrow n + e^+ \\
         \bar{\nu}_{\mu}+ p \longrightarrow n + \mu^+ \nonumber
   \end{eqnarray}

The differential cross section for QEL is
given as follows \cite{r6}.\\

    \begin{eqnarray}
         \frac{{\rm d}\sigma}{{\rm d}Q^2} = 
         \frac{G_F^2{\rm cos}^2 \theta_C}{8\pi E_{\nu}^2}
         \Biggl\{ A(Q^2) \pm B(Q^2) \biggl[ \frac{s-u}{M^2} \biggr]
         +C(Q^2) \biggl[ \frac{s-u}{M^2} \biggr]^2 \Biggr\},
    \end{eqnarray}

\noindent where

    \begin{eqnarray*}
      A(Q^2) &=& \frac{Q^2}{4}\Biggl[ f^2_1\biggl( \frac{Q^2}{M^2}-4 \biggr)
      + f_1f_2 \frac{4Q^2}{M^2}  + f_2^2\biggl( \frac{Q^2}{M^2}
      -\frac{Q^4}{4M^4} \biggr) + g_1^2\biggl( 4+\frac{Q^2}{M^2} \biggl)
       \Biggr], \\
      B(Q^2) &=& (f_1+f_2)g_1Q^2, \\
      C(Q^2) &=& \frac{M^2}{4}\biggl( f^2_1+f^2_2\frac{Q^2}{4M^2}+g_1^2 \biggr).
    \end{eqnarray*}

\noindent The signs $+$ and $-$ refer to $\nu_{\mu(e)}$ and $\bar{\nu}_{\mu(e)}$
for charged current (c.c.) interactions, respectively.  The $Q^2$ denotes
the four momentum transfer between the incident neutrino and the charged
lepton. Details of other symbols are given in \cite{r6}.

The relation among $Q^2$ , $E_{\nu}$, the energy of the incident
neutrino, $E_{\ell}$, the energy of the emitted charged lepton
(muon or electron or their anti-particles) 
and $\theta_{\rm s}$, the scattering
angle of the charged lepton, is given as

      \begin{equation}
         Q^2 = 2E_{\nu}E_{\ell}(1-{\rm cos}\theta_{\rm s}).
      \end{equation}

\noindent Also, the energy of the charged lepton is given by

      \begin{equation}
         E_{\ell} = E_{\nu} - \frac{Q^2}{2M}.
      \end{equation}

\begin{figure}
\vspace{-40mm}
\hspace*{-15mm}
\includegraphics[scale=0.6,angle=90]{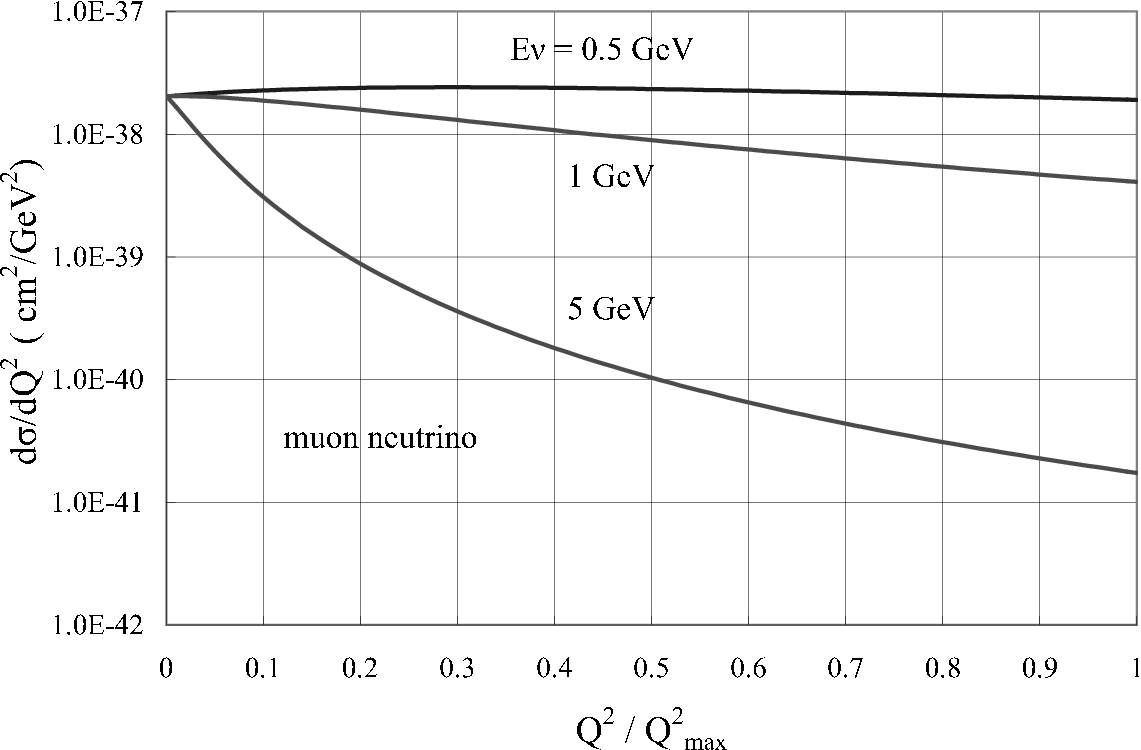}
\vspace{-5mm}
\caption{\label{fig:1} Differential cross section for QEL for muon neutrinos with  different incident energies,
 0.5 GeV, 1GeV and 5 GeV.}

\hspace*{-15mm}
\includegraphics[scale=0.6,angle=90]{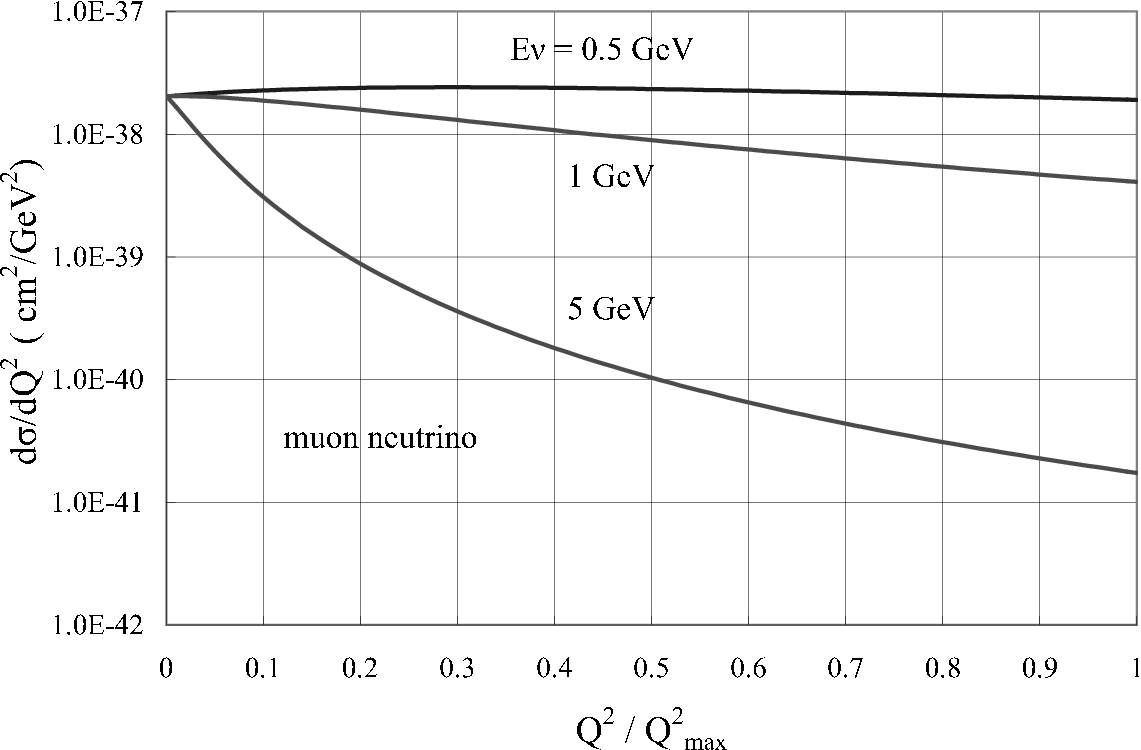}
\vspace{-5mm}
\caption{\label{fig:2} Differential cross section for QEL for anti-muon 
neutrinos with the same incident energies as in Fig. 1.}
\end{figure} 

\begin{figure}
\vspace{-40mm}
\hspace*{-15mm}
\includegraphics[scale=0.6,angle=90]{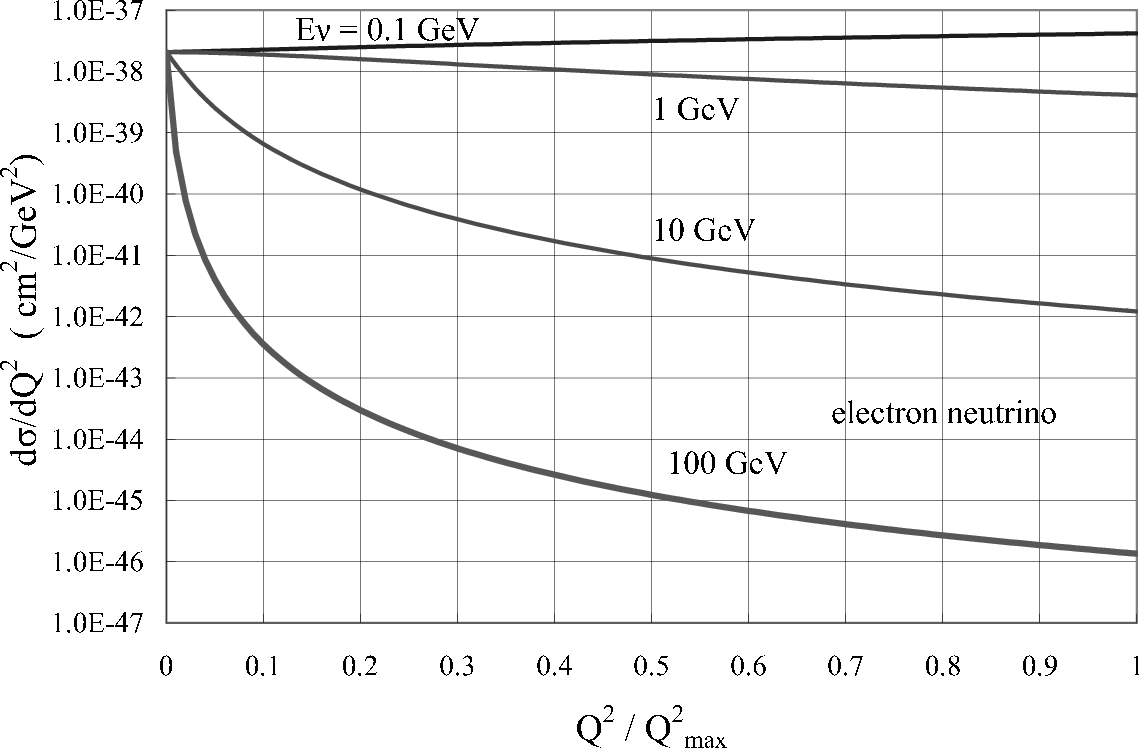}
\vspace{-5mm}
\caption{\label{fig:3} Differential cross section for QEL for electron 
neutrinos with  diffrent incident energies, 0.1 GeV, 1 GeV, 10 GeV
 and 100 GeV.}

\hspace*{-15mm}
\includegraphics[scale=0.6,angle=90]{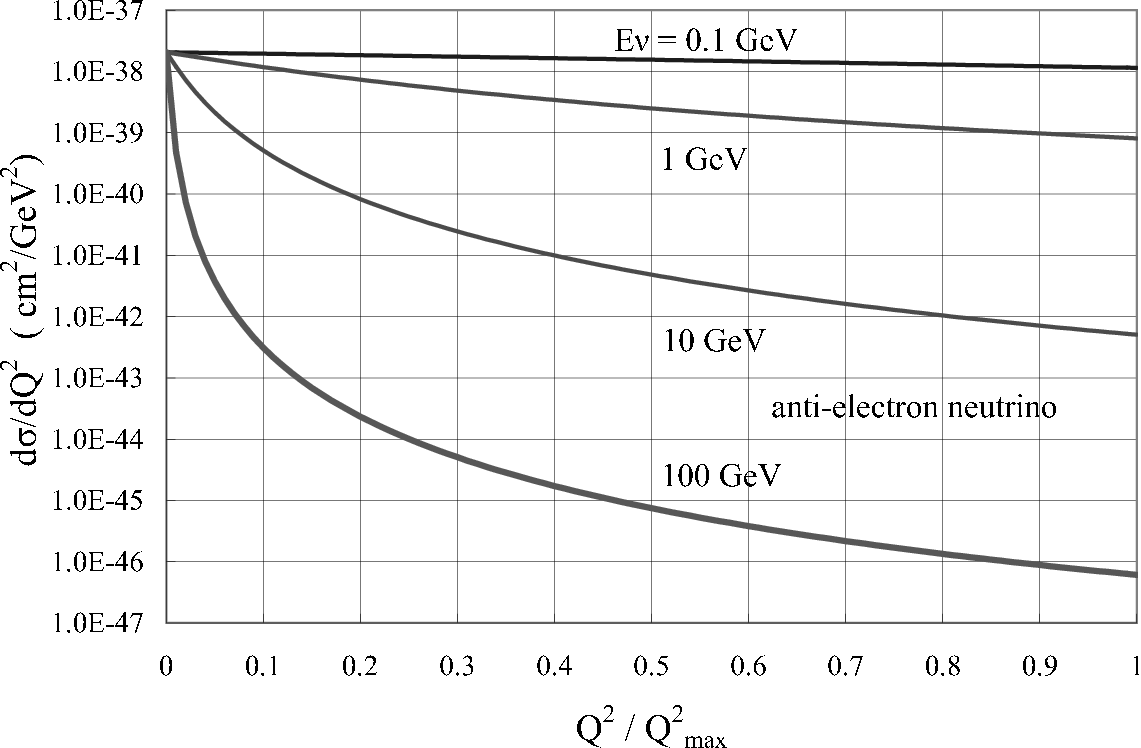}
\vspace{-5mm}
\caption{\label{fig:4} Differential cross section for QEL for anti-electron
 neutrinos with the same incident energies as in Fig. 3.}
\end{figure}


In Figs. 1 to 4, we give the differential cross sections for different
charged leptons and for different incident neutrino energies. It is
clear from these figures that the cross sections of the neutrinos are
larger than those of anti-neutrinos in the lower energy region, say,
$\sim$0.1~GeV to $\sim$10~GeV and that their difference are negligible in higher
energy region, say, beyond $\sim$10~GeV. The differences between (anti-)muon
and (anti-)electron are negligible except in the lower energy region,
below $\sim$0.1~GeV.

In Fig. 5, we give the total cross sections for muon neutrinos and
anti-muon neutrinos. It can be seen that the differences between them
are rather large in the lower energy region, say, $\sim$0.1~GeV to $\sim$10~GeV, but
that the cross sections have similar values above $\sim$10~GeV. 
The corresponding total cross sections for electron and
positron are almost the same as those of the muon neutrinos and anti-muon
neutrinos, respectively.

\begin{figure}
\hspace*{-20mm}
\includegraphics[scale=0.58,angle=90]{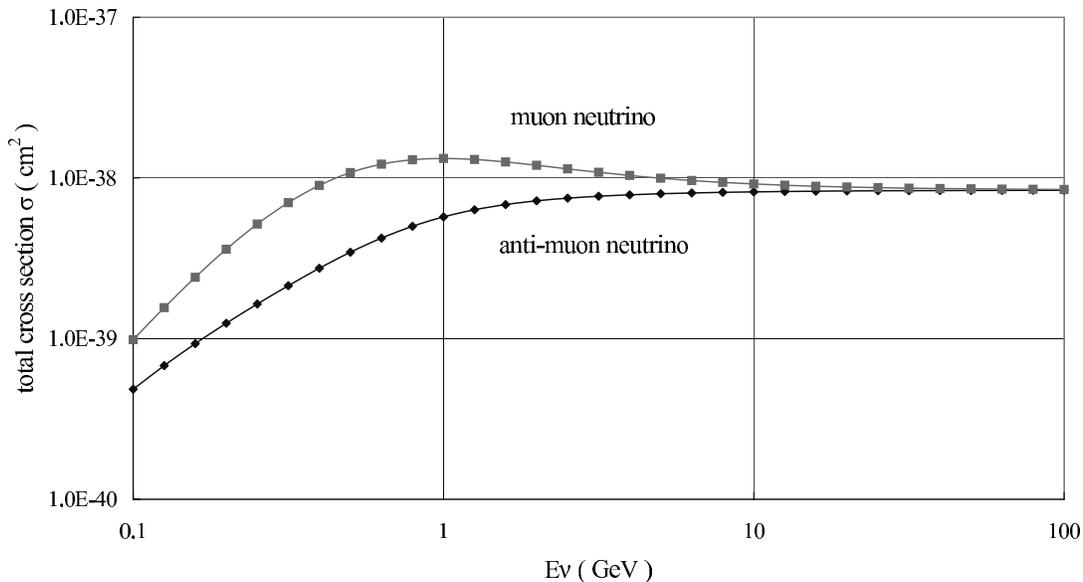}
\vspace{-5mm}
\caption{\label{fig:5} Total cross sections of QEL for muon neutrino and 
 anti-muon neutrino as a function of the incident neutrino energy.}
\end{figure}

Now, let us examine the magnitude of the scattering angle of the
charged lepton in a quantitative way, as this plays a decisive role in
determining the accuracy of the direction of the incident neutrino,
which is directly related to the reliability of the zenith angle
distribution of both {\it Fully Contained Events} and {\it Partially
Contained Events} in SK.

It can be seen from Eqs. (3) and (4) that there is a relation between
the energy $E_{\ell}$ of the charged lepton and its scattering angle $\theta_{\rm s}$ for a given
incident neutrino energy $E_{\nu}$.  Figure 6 shows this relation for muon, from
which we can easily understand that the scattering angle $\theta_{\rm s}$ of the
charged lepton ( muon here ) cannot be neglected.  For a quantitative examination of
the scattering angle, we construct the distribution function for ${\theta_{\rm s}}$ of the charged lepton from Eqs. (2) to (4) by using the Monte Carlo method.
\footnote {The distribution functions for the scattering angle can be more 
easily obtained by the numerical method than done by the Monte
 Carlo method. The reason why we obtain them by the Monte Carlo method
 is to apply this Monte Carlo simulation procedure to another purpose 
 by which we get the essential part of this paper. The distribution 
 function for scattering angle in the inelastic scattering of the 
 neutrino interaction and other quantities had been obtained by Kobayakawa {\it et al.}
 See \cite{r7}.
 }

 Figure 7 gives the distribution function for $\theta_{\rm s}$ of the muon produced in the muon neutrino interaction. It can be seen 
that the muons
produced from lower energy neutrinos are scattered over wider angles
and that a considerable part of them are scattered even in backward directions. 
Similar results are obtained for anti-muon neutrinos,
electron neutrinos and anti-electron neutrinos.

\begin{figure}
\vspace{-40mm}
\hspace*{-15mm}
\includegraphics[scale=0.55,angle=90]{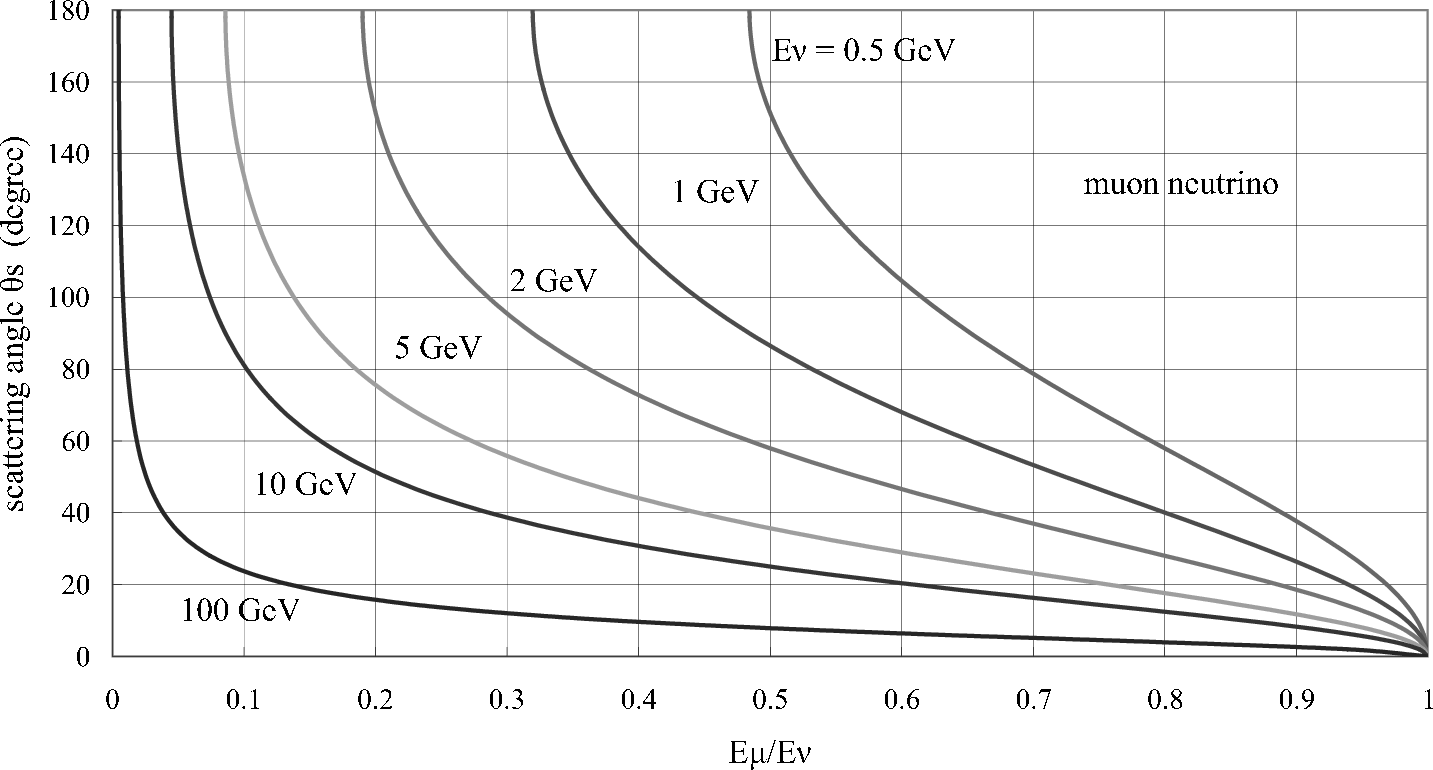}
\vspace{-5mm}
\caption{\label{fig:6} Relation between the energy of the muon and its 
scattering angle for different incident muon neutrino energies,
 0.5~GeV, 1~GeV, 2~GeV, 5~GeV, 10~GeV and 100~GeV.}

\hspace*{-10mm}
\includegraphics[scale=0.6,angle=270]{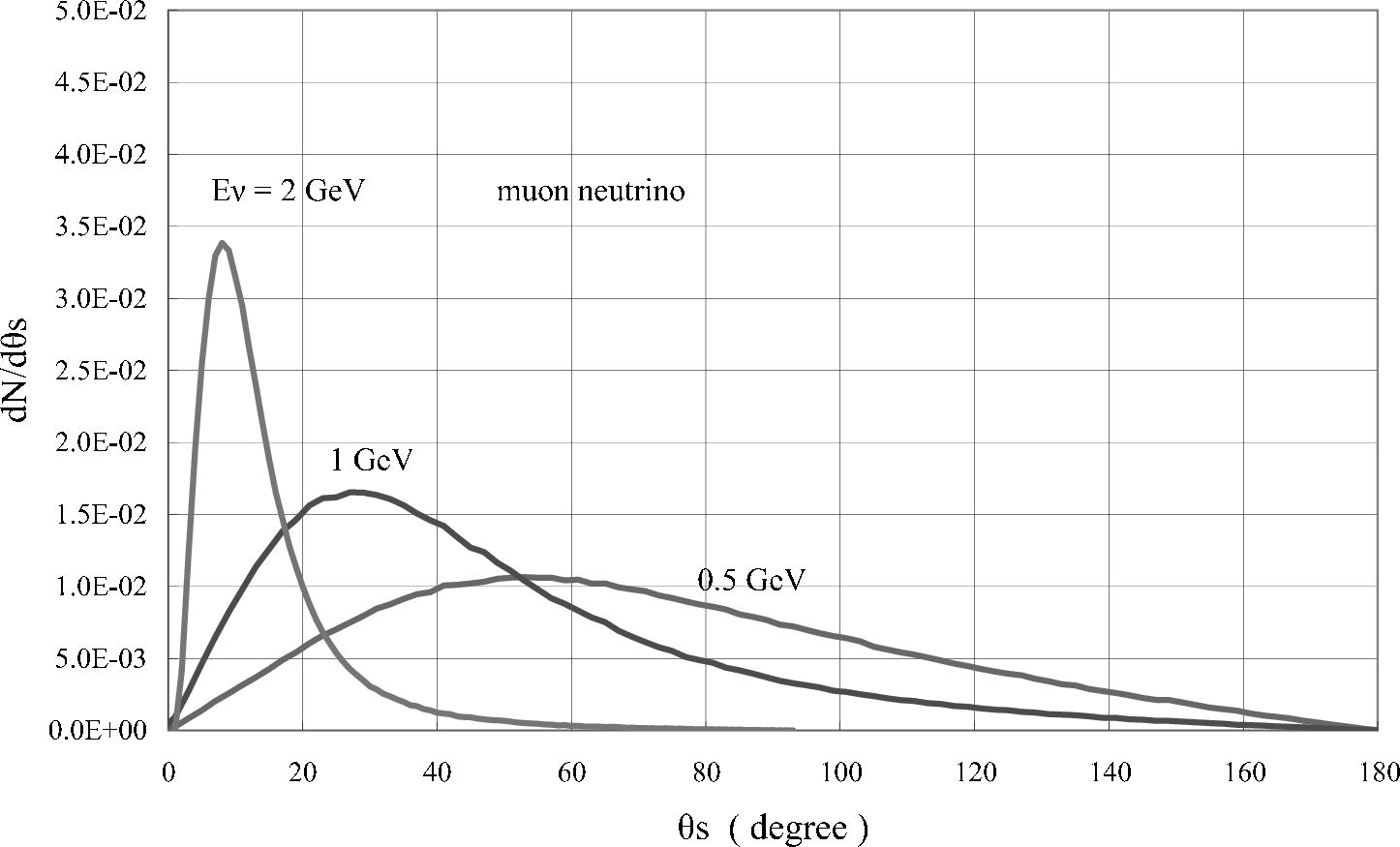}
\vspace{-5mm}
\caption{\label{fig:7} Distribution functions for the scattering angle of 
the muon for muon-neutrino with incident energies, 0.5 GeV, 1.0 GeV and 2 GeV. Each curve is obtained by the Monte Carlo 
method (one million sampling per each curve). }
\end{figure}

Also, in a similar manner, we obtain not only the distribution function
for the scattering angle of the charged leptons, but also their
average values $<\theta_{\rm s}>$ and their standard deviations $\sigma_{\rm s}$. Table~1 shows them
 for muon neutrinos, anti-muon
neutrinos, electron neutrinos and anti-electron neutrinos.  In the SK
analysis, it is assumed that the scattering angle of the charged
particle is zero \cite{r3, r4}, but Table~1 demonstrates that
such an assumption does not hold at all in both {\it Fully Contained
Events} and {\it Partially Contained Events}.

Therefore, it is surely concluded that the zenith angle distribution obtained by SK for {\it
 Fully Contained Events} and {\it Partially Contained Events} are almost 
 unreliable, which will be shown later more concretely.

\begin{center}
\begin{table}
\caption{\label{tab:table1} The average values $<\theta_{\rm s}>$ for scattering angle of the 
emitted charged leptons and their standard deviations $\sigma_{\rm s}$ for various primary 
neutrino energies $E_{\nu}$.}
\vspace{5mm}
\begin{tabular}{|c|c|c|c|c|c|}
\hline
&&&&&\\
$E_{\nu}$ (Gev)&angle&$\nu_{\mu}$&$\bar{\nu_{\mu}}$&$\nu_e$&$\bar{\nu_e}$ \\
&(degree)&&&&\\
\hline
0.2&$<\theta_\mathrm{s}>$&~~ 89.86 ~~&~~ 67.29 ~~&~~ 89.74 ~~&~~ 67.47 ~~\\
\cline{2-6}
   & $\sigma_\mathrm{s}$ & 38.63 & 36.39 & 38.65 & 36.45 \\
\hline
0.5&$<\theta_\mathrm{s}>$& 72.17 & 50.71 & 72.12 & 50.78 \\
\cline{2-6}
   & $\sigma_\mathrm{s}$ & 37.08 & 32.79 & 37.08 & 32.82 \\
\hline
1  &$<\theta_\mathrm{s}>$& 48.44 & 36.00 & 48.42 & 36.01 \\
\cline{2-6}
   & $\sigma_\mathrm{s}$ & 32.07 & 27.05 & 32.06 & 27.05 \\
\hline
2  &$<\theta_\mathrm{s}>$& 25.84 & 20.20 & 25.84 & 20.20 \\
\cline{2-6}
   & $\sigma_\mathrm{s}$ & 21.40 & 17.04 & 21.40 & 17.04 \\
\hline
5  &$<\theta_\mathrm{s}>$&  8.84 &  7.87 &  8.84 &  7.87 \\
\cline{2-6}
   & $\sigma_\mathrm{s}$ &  8.01 &  7.33 &  8.01 &  7.33 \\
\hline
10 &$<\theta_\mathrm{s}>$&  4.14 &  3.82 &  4.14 &  3.82 \\
\cline{2-6}
   & $\sigma_\mathrm{s}$ &  3.71 &  3.22 &  3.71 &  3.22 \\
\hline
100&$<\theta_\mathrm{s}>$&  0.38 &  0.39 &  0.38 &  0.39 \\
\cline{2-6}
   & $\sigma_\mathrm{s}$ &  0.23 &  0.24 &  0.23 &  0.24 \\
\hline
\end{tabular}
\end{table}
\end{center}

\section{Influence of Azimuthal Angle of Quasi Elastic Scattering 
over the Zenith Angle of the Fully Contained Events
}

   For three incident cases(vertical, horizontal and diagonal), Fig. 8 gives a schematic representation of the relationship among the zenith angle of the incident neutrino, $\theta_{\nu}$, the scattering angle of the charged lepton, $\theta_{\rm s}$, and the azimuthal angle of the charged lepton, $\phi$. In this paper, we measure $\theta_{\nu}$ from the vertical upward direction.

  From Fig. 8-a, it can been seen that the zenith angle $\theta$ of the charged lepton is not influenced by its $\phi$ in the vertical incidence $(\theta_{\nu}=0^{\rm o})$. From Fig. 8-b, however, it is obvious that the angle $\phi$ influences on its zenith angle greatly in the case of horizontal incidence of the neutrino $(\theta_{\nu}=90^{\rm o})$. Namely, half of the charged leptons are recognized as upward going, while the other half is classified as downward going. The intermediate case (diagonal incidence of $\theta_{\nu}=43^{\rm o}$) between the above two cases is shown in Fig. 8-c.
   As discussed above, we must take account of the effect due to the azimuthal angle $\phi$ in the analysis of both {\it Fully Contained Events} and {\it Partially Contained Events}.
    Figure 9 shows the relation between direction cosines of the incident neutrinos, $(\ell, m, n )$, and those of the charged lepton, $(\ell_{\rm r}, m_{\rm r}, n_{\rm r})$, for certain $\theta_{\rm s}$ and $\phi$. 
    Then, we have the following expression:

\begin{figure}
\includegraphics[scale=0.6]{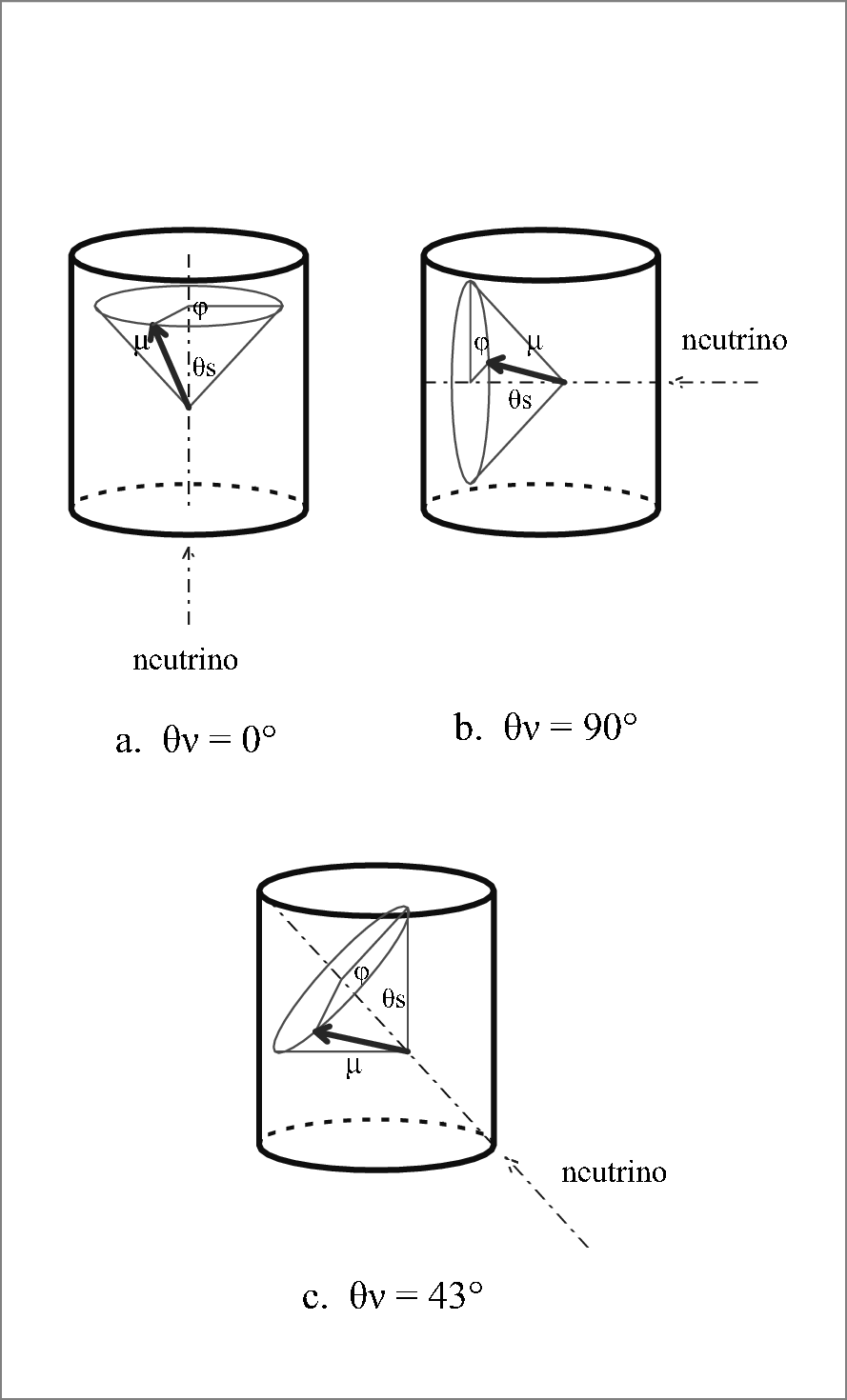}
\caption{\label{fig:8} Schematic view of the zenith angles of the charged
 muons for diffrent zenith angles of the incident neutrinos, focusing on
  their azimuthal angles.}
\end{figure}

\begin{figure}
\includegraphics[scale=0.6]{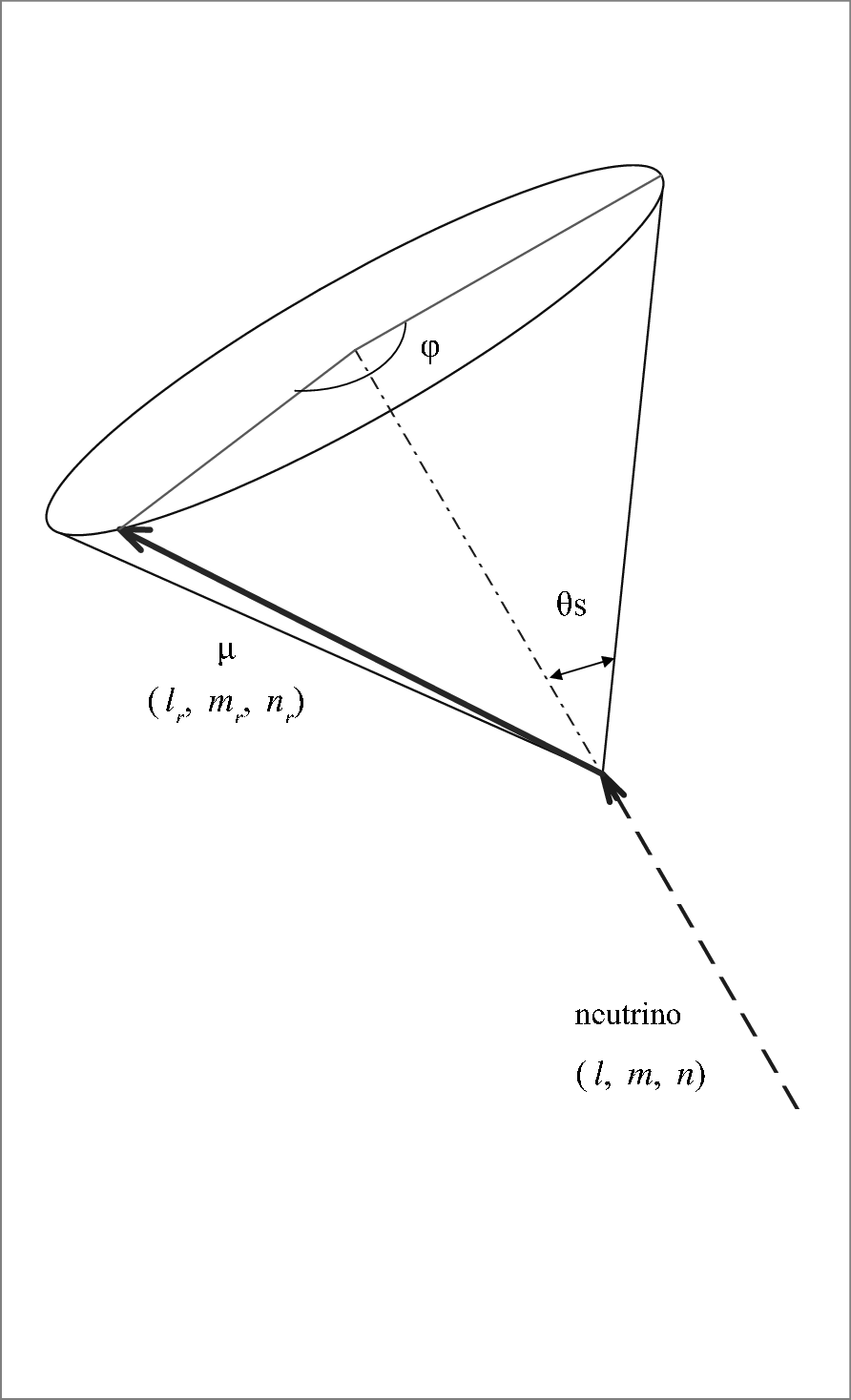}
\caption{\label{fig:9} The relation between the direction cosines of the
 incident neutrino and those of the emitted charged lepton.}
\end{figure}

\begin{equation}
\left(
         \begin{array}{c}
             \ell_{\rm r} \\
             m_{\rm r} \\
             n_{\rm r}
         \end{array}
       \right)
           =
       \left(
         \begin{array}{ccc}
            \displaystyle \frac{\ell n}{\sqrt{\ell^2+m^2}} & -\displaystyle 
            \frac{m}{\sqrt{\ell^2+m^2}} & \ell \\
            \displaystyle \frac{mn}{\sqrt{\ell^2+m^2}} & \displaystyle 
            \frac{\ell}{\sqrt{\ell^2+m^2}} & m \\
                        -\sqrt{\ell^2+m^2} & 0 & n
         \end{array}
       \right)
       \left(
          \begin{array}{c}
            {\rm sin}\theta_{\rm s}{\rm cos}\phi \\
            {\rm sin}\theta_{\rm s}{\rm sin}\phi \\
            {\rm cos}\theta_{\rm s},
          \end{array}
       \right),
\end{equation}

\noindent where $n={\rm cos}\theta_{\nu}$, and $n_{\rm r}={\rm cos}\theta$. Here, $\theta$ is the zenith angle of the charged lepton. 

Using Eq. (5), we carry out a Monte Carlo calculation to examine the
influence of $\phi$ on $\theta$ for the charged lepton. The Monte Carlo procedure for the determination of the real $\theta$ of the charged lepton whose parent (anti-)neutrino has 
fixed $\theta_{\nu}$ and $E_{\nu}$ involves the following steps:\\

\noindent
1. We extract $Q^2$ from the probability function for the
differential cross section with a given $E_{\nu}$ (Eq. (2)) by the
random sampling.\\
2. We obtain $E_{\ell}$ from Eq. (4).\\
3. We obtain $\theta_{\rm s}$ from Eq. (3).\\
4. We decide $\phi$, which is obtained from

  \begin{equation}
       \phi = 2\pi\xi.
  \end{equation}

\noindent Here, $\xi$ is a uniform random number between 0 and 1. \\
5. We obtain $(\ell_r, m_r, n_r)$ by using Eq. (5) . The
$n_r$ is the cosine of the zenith angle of the charged lepton which
should be contrasted to $n$, that of the incident
neutrino.
In the SK analysis, it is assumed that

    \begin{equation}
        (\ell_{\rm r}, m_{\rm r}, n_{\rm r}) =(\ell, m, n).
    \end{equation}
\\

We examine $\theta$, the zenith angles of the charged muons, for three typical
zenith angles of the incident neutrino with $E_{\nu}=1$ GeV, namely, $\cos\theta_{\nu}=1$ 
(vertically upward), $\cos\theta_{\nu}=0.731$, i.e., $\theta_{\nu} =43^\circ$ 
(diagonal in SK), and $\cos\theta_{\nu}=0$ (horizontal).\\

  \par
A : vertically upward incident neutrino events\\
In this case, as easily understood from Fig. 8-a, we can skip 
the step 4 mentioned above, because the change of $\phi$ of the charged muon has no influence over its zenith angle $\theta$.

Here, although the direction of the charged muon approximately
retains the primary direction of the incident neutrino at higher
energies, say, above $\sim$20~GeV, it deviates from the
primary direction (vertical), even to backward directions, at lower
energies, say, below $\sim$1~GeV (see also Fig.~7).

\begin{figure}
\vspace{-40mm}
\hspace*{10mm}
\includegraphics[scale=0.4,angle=270]{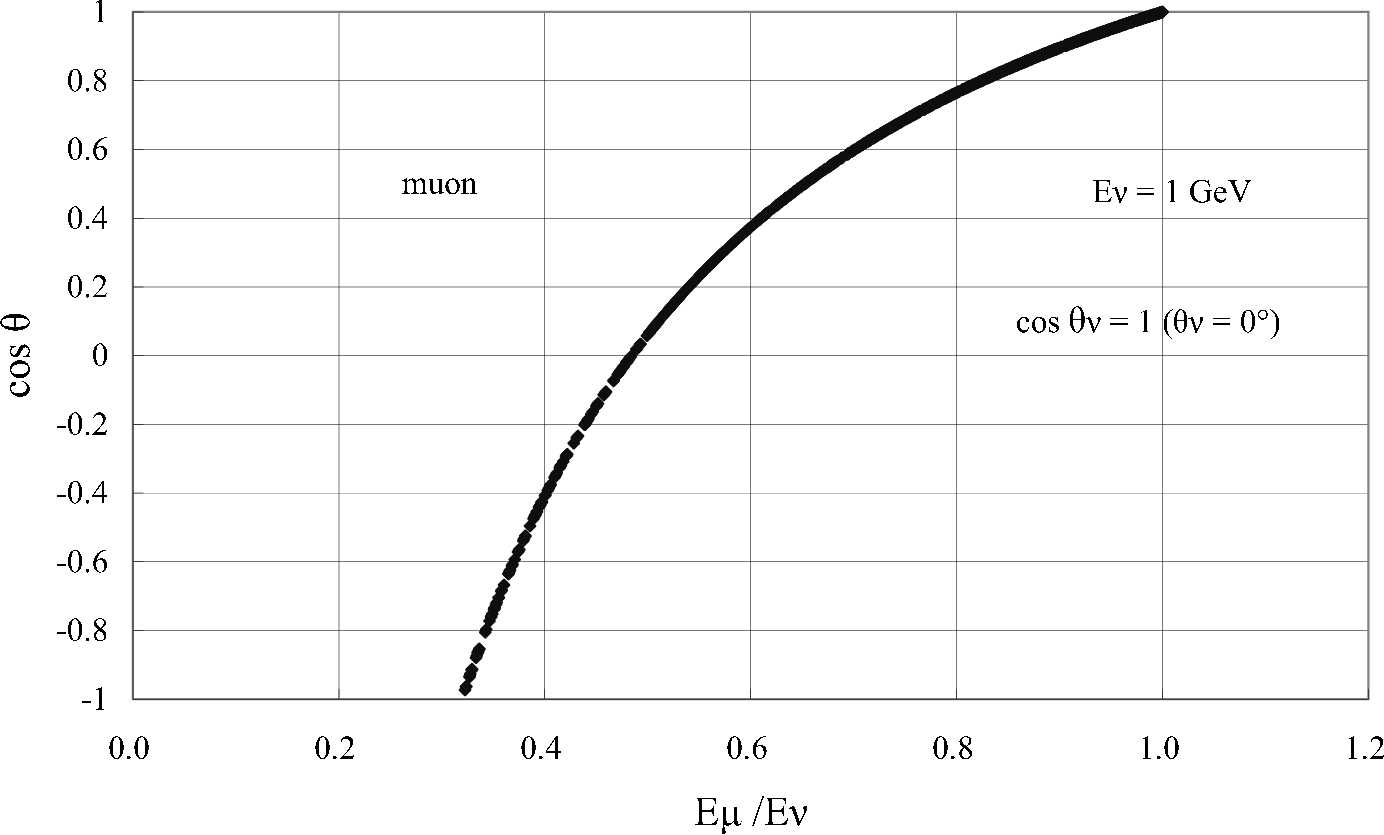}
\vspace{-5mm}
\caption{\label{fig:10} The scatter plot between the fractional energies of 
the produced muons and their zenith angles for vertically incident muon neutrinos with 1~GeV.  The sampling number is 1000.}

\hspace*{10mm}
\includegraphics[scale=0.4,angle=270]{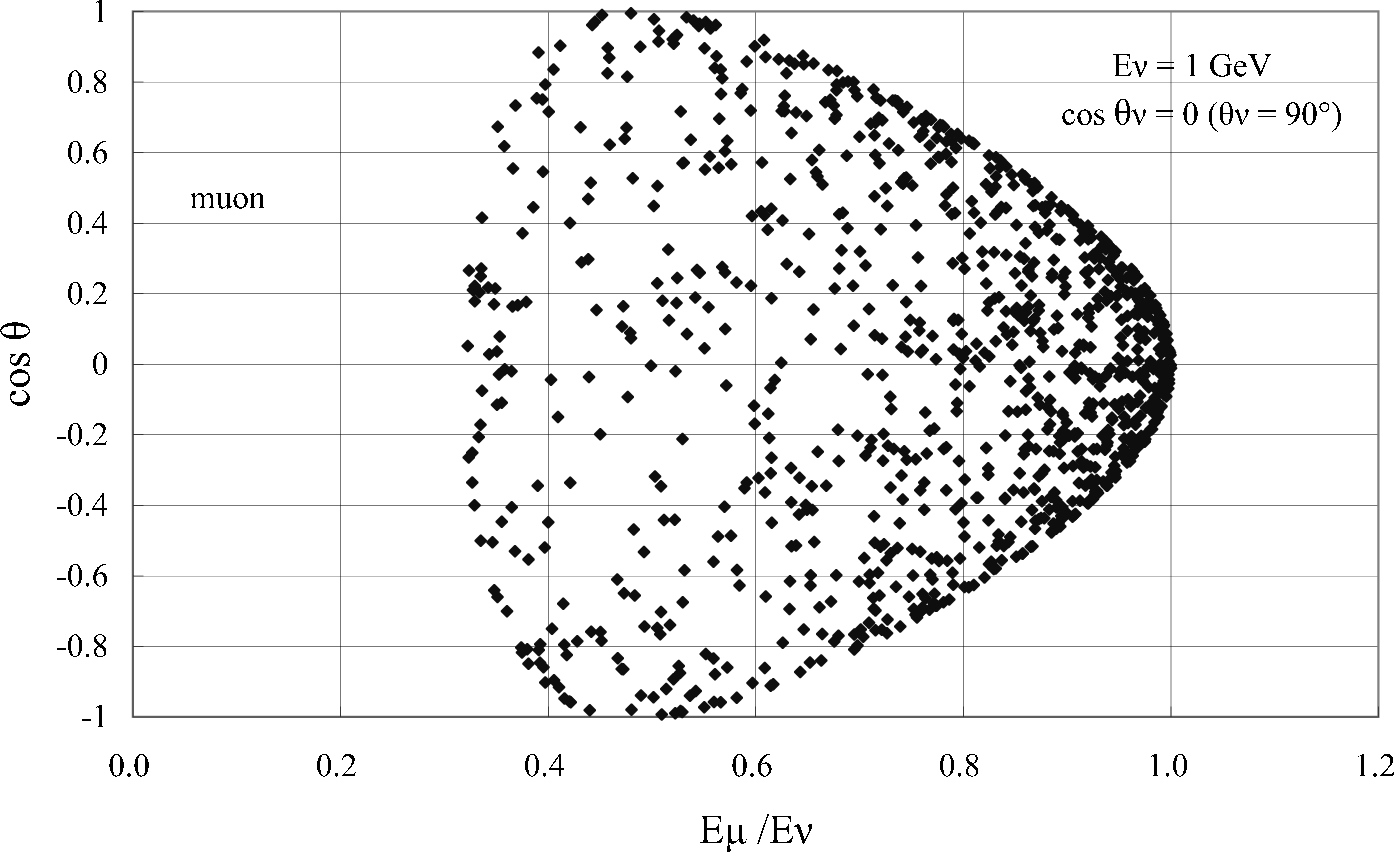}
\vspace{-5mm}
\caption{\label{fig:11} The scatter plot between the fractional energies of 
the produced muons and their zenith angles for horizontally incident muon neutrinos with 1~GeV. The sampling number is 1000.}

\hspace*{10mm}
\includegraphics[scale=0.4,angle=270]{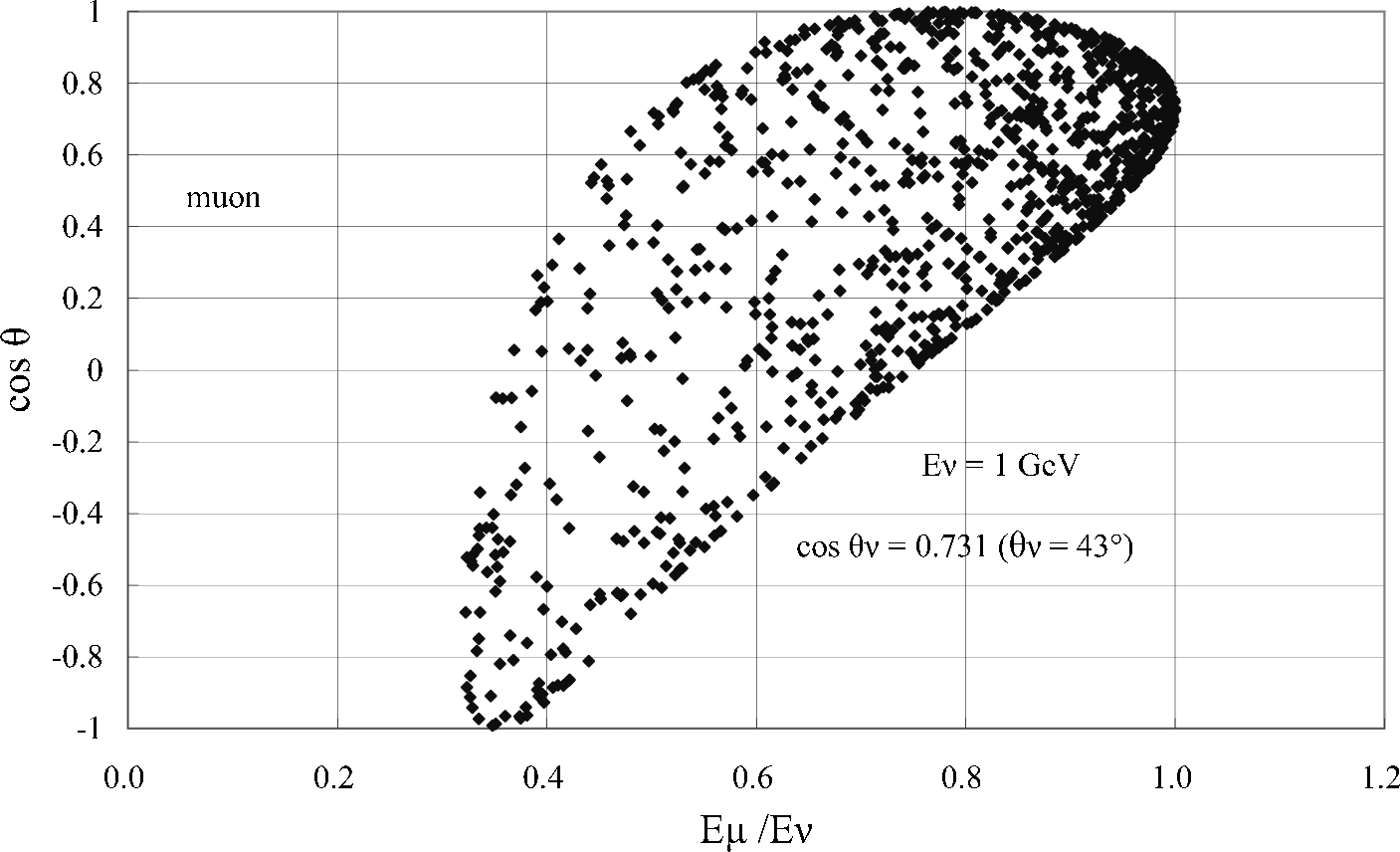}
\vspace{-5mm}
\caption{\label{fig:12} The scatter plot between the fractional energies of 
the produced muons and their zenith angles for diagonally incident muon neutrinos with 1~GeV. The sampling number is 1000.}
\end{figure} 

In Fig. 10, we show the scatter plot between $E_{\mu}$, the
emitted energy of the charged muon and $\cos\theta$, the cosine of
the zenith angle of the charged muon for the vertically incident neutrino.
In this case, the relation between $\theta$ and $E_{\nu}$ is unique because of  independence of
the azimuthal angle. The reason is as follows: For a given $Q^2$ (see Eq.(2))
 we obtain the energy of the charged lepton and its scattering angle uniquely due to the two body kinematics and the zenith 
  angle of the charged lepton are never influenced from their azimuthal 
  angle, because $\theta$ is measured from the vertical direction
, parallel to the axis of the detector.\\

  \par
B: horizontally incident neutrino events\\
This case shows a great contrast to the vertical case. In Fig. 11, we give the
scatter plot between $E_{\nu}$ and ${\rm cos}\theta$.In this case,
the influence of $\phi$ over $\theta$ is the
strongest amomg the three, because this influence appears through Eq.(5) as strongly as possible. The ${\rm cos}\theta$ has the widest distribution, even if the charged muon has the same energy.  As $\theta$ is symmetrical due to its azimuthal symmetry, the scattered muons have almost lost the direction of primary neutrinos at lower energies, say, $\sim$1~GeV and show a uniform distribution between upward and downward. It should be noted that the scatter plots are distributed
symmetrically around $\cos\theta=0$, reflecting the azimuthal
symmetry.\\

\par
C: diagonally upward incident neutrino events\\
In Fig. 12, we show the similar scatter plot for the diagonal case,
which shows the intermediate situation between Figs. 10 and 11, as expected.  From Fig. 12 for $\theta_{\nu}=43^{\rm o}$, it is still apparent that $\phi$ influences
$\theta$ in a considerable degree, even if the energies of the
charged muon are the same.\\

\par

We can get the similar scatter plots for electrons to those for muons.

We sum the events concerned over $E_{\ell}$ for a given $\rm cos\theta$ in Figs 10 to 12 and show the results of the zenith angle disributin in Figs 13 to 15. In
Fig. 13, the neutrino enters vertically. In this case, even if
we define $\cos\theta= 0.8\sim 1.0$ as ``vertical'', about one half of  the real events are not
recognized as vertical.  In Fig. 14 we give the zenith angle distribution  for the horizontal($\cos\theta_{\nu}=0$).  Comparing it with Fig. 13, it is easily understood
that the direction of the primary neutrino here has been
almost completely lost, namely, $n_{\rm r}$ is very different from
$n$. Thus, the SK assumption $(n_{\rm r} = n)$ clearly does not
hold in this energy region. In Fig. 15, we show the intermediate
state between Figs. 13 and 14. The tendencies in Figs.~13 to 15
for muon neutrinos are quite similar to those for electron neutrinos.

Furthermore, the dependence of ${\rm cos}\theta$ on ${\rm cos}\theta_{\nu}$ becomes more sensitive at lower energies, say, below $\sim$0.2~GeV. 
Also, they are a little weaker for anti-neutrinos than for neutrinos, 
as can be inferred from the cross sections in Figs.~1 to 4 and Fig.~5.

  It should be, here, emphasized that the azimuthal angle of the charged lepton greatly influences discrimination between {\it Fully Contained Events} and
   {\it Partially Contained Events}, which is strongly dependent on the generation points of the neutrino events concerned. The estimation of 
   this discrimination is very important for getting 
    {\it Fully Contained Events}, because the correct analysis of 
    {\it Fully Contained Events} only in the SK experiment makes it possible
     to lead less ambigious  conclusion on neutrino oscillation.         @

\begin{figure}
\vspace{-30mm}
\hspace*{10mm}
\includegraphics[scale=0.4,angle=90]{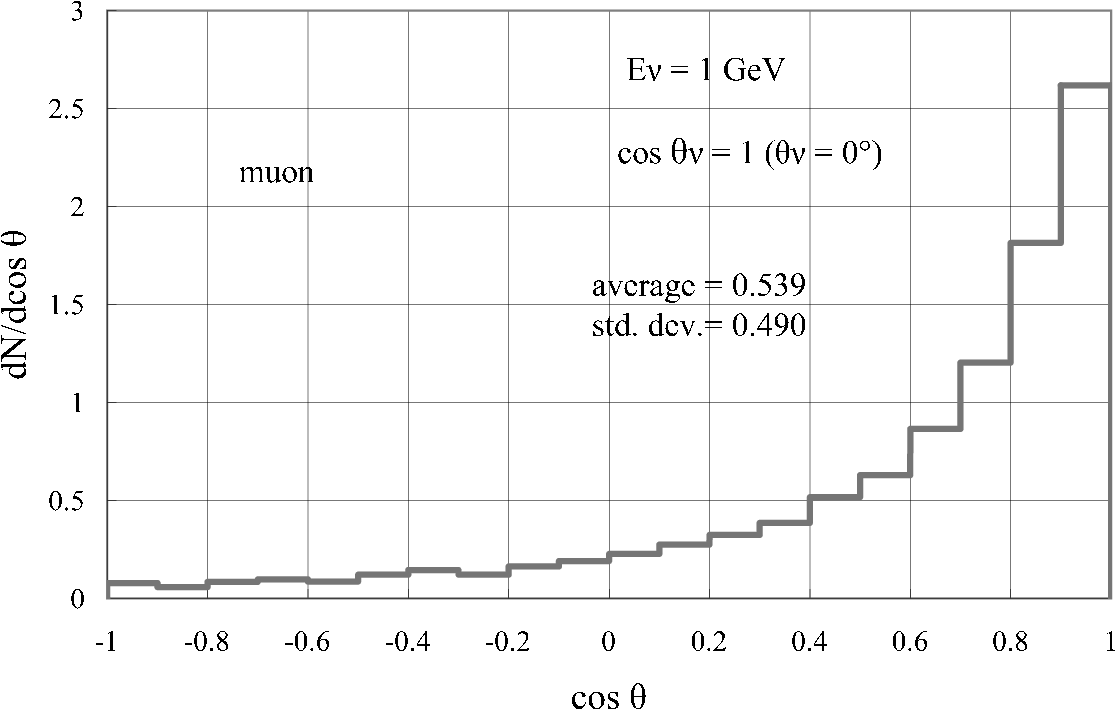}
\vspace{-5mm}
\caption{\label{fig:13} Zenith angle distribution of the muon for the vertically incident muon neutrino with 1 GeV. The sampling 
number is 10000.}

\hspace*{10mm}
\includegraphics[scale=0.4,angle=90]{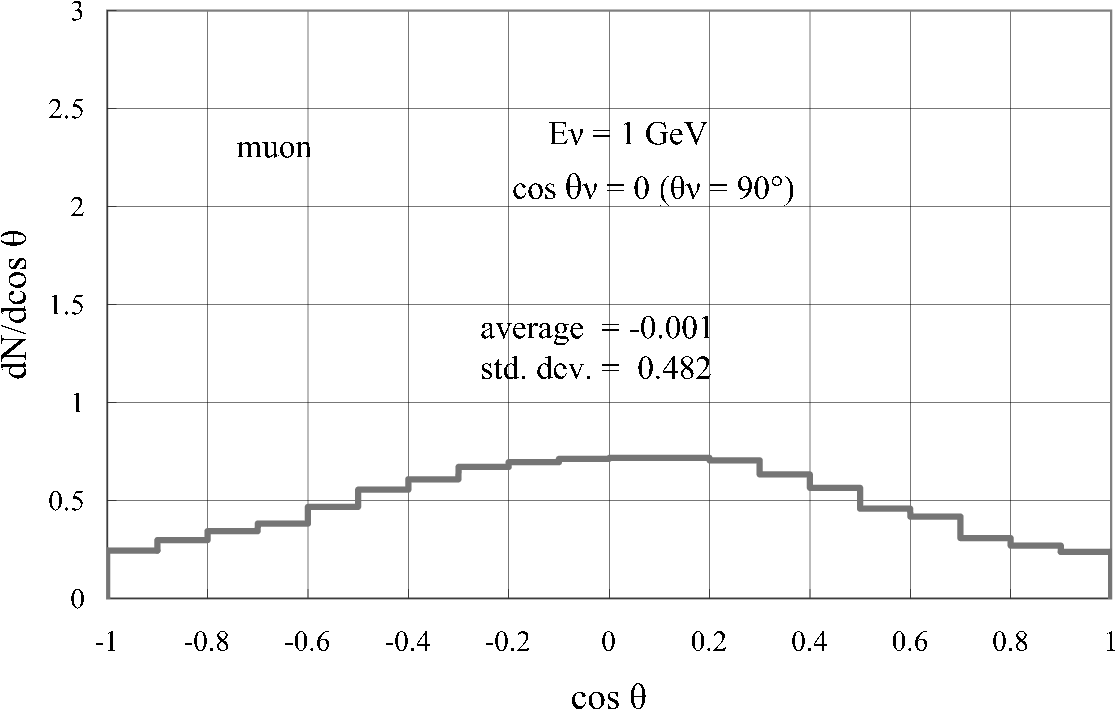}
\vspace{-5mm}
\caption{\label{fig:14} Zenith angle distribution of the muon 
for the horizontally incident muon neutrino with 1 GeV.
 The sampling number is 10000.}

\hspace*{10mm}
\includegraphics[scale=0.4,angle=90]{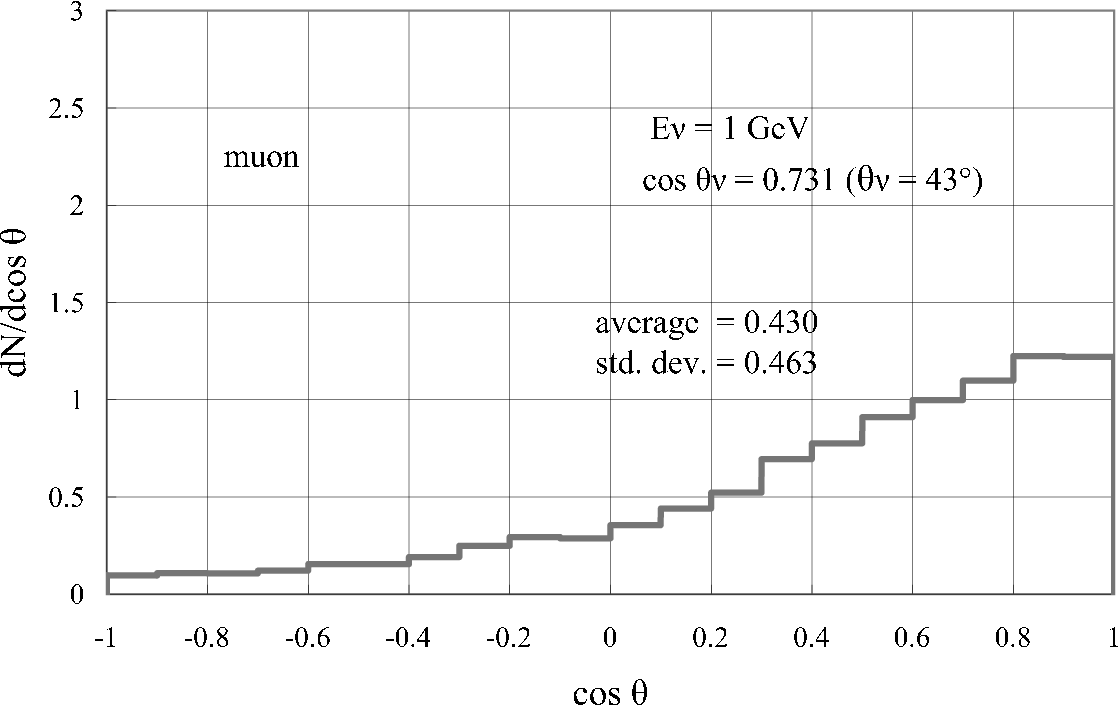}
\vspace{-5mm}
\caption{\label{fig:15} Zenith angle distribution of the muon for the diagonally incident muon 
neutrino with 1 GeV. The sampling number is 10000.}
\end{figure}

\section{The Zenith Angle of Fully Contained Events 
Taking Account of the Overall Neutrino Spectrum}

   In the previous section, we obtain the zenith angle distribution of both
{\it Fully Contained Events} and {\it Partially Contained Events} for
a given energy of the incident neutrino, taking account of the
effect of the azimuthal angle for a given zenith angle of the incident 
neutrino. However, in
order to examine the effect of $\theta_{\rm s}$ and $\phi$ of the charged lepton in QEL on the direction of primary neutrinos, we must consider the overall
neutrino spectrum, not a fixed neutrino energy.\\

The concrete procedures for a selected event with a given $\theta_{\nu}$ of the incident neutrino are summarized again as follows:\\

Procedure A\\
\noindent
We formulate the differential neutrino interaction probability
functions for muon (electron) and anti-muon (anti-electron) in which the
(anti-)neutrino differential energy spectra are combined with the total cross
sections of the corresponding neutrino concerned for QEL. In such a calculation,the differential neutrino spectra at Kamioka site which cover from 0.1~GeV to
100~GeV obtained by Fiorentini {\it et al.}\cite{r8} are utilized. \\

Procedure B\\
\noindent
The determination as to whether the (anti-)neutrino concerned belongs
to either neutrino or to anti-neutrino is done by random sampling
according to the probability functions constructed from Procedure A.\\

Procedure C\\
\noindent
We randomly sample the energy of the (anti-)neutrino concerned according to
the probability function constructed from the Procedure A.\\

Procedure D\\
\noindent
We randomly sample $Q^2$  from  Eq. (2) whose neutrino energy is determined
by Procedure C.\\

Procedure E\\
\noindent
We decide the energy of the charged lepton concerned, $E_{\ell}$, for the $Q^2$ from Eq. (4).\\

Procedure F\\
\noindent
We decide the scattering angle of the charged lepton concerned, 
${\theta_{\rm s}}$,
from Eq. (3).\\

Procedure G\\
\noindent
We randomly sample the azimuthal angle of the charged lepton concerned,
$\phi$, from Eq. (6).\\
 
Procedure H\\
\noindent
We decide the direction cosines of the charged lepton concerned by using 
Eq. (5), being accompanied by Procedures F and G.\\   

 We repeat Procedures B to H until we reach the desired trial
number. \\

\begin{figure}
\vspace{-37mm}
\hspace*{10mm}
\includegraphics[scale=0.4,angle=90]{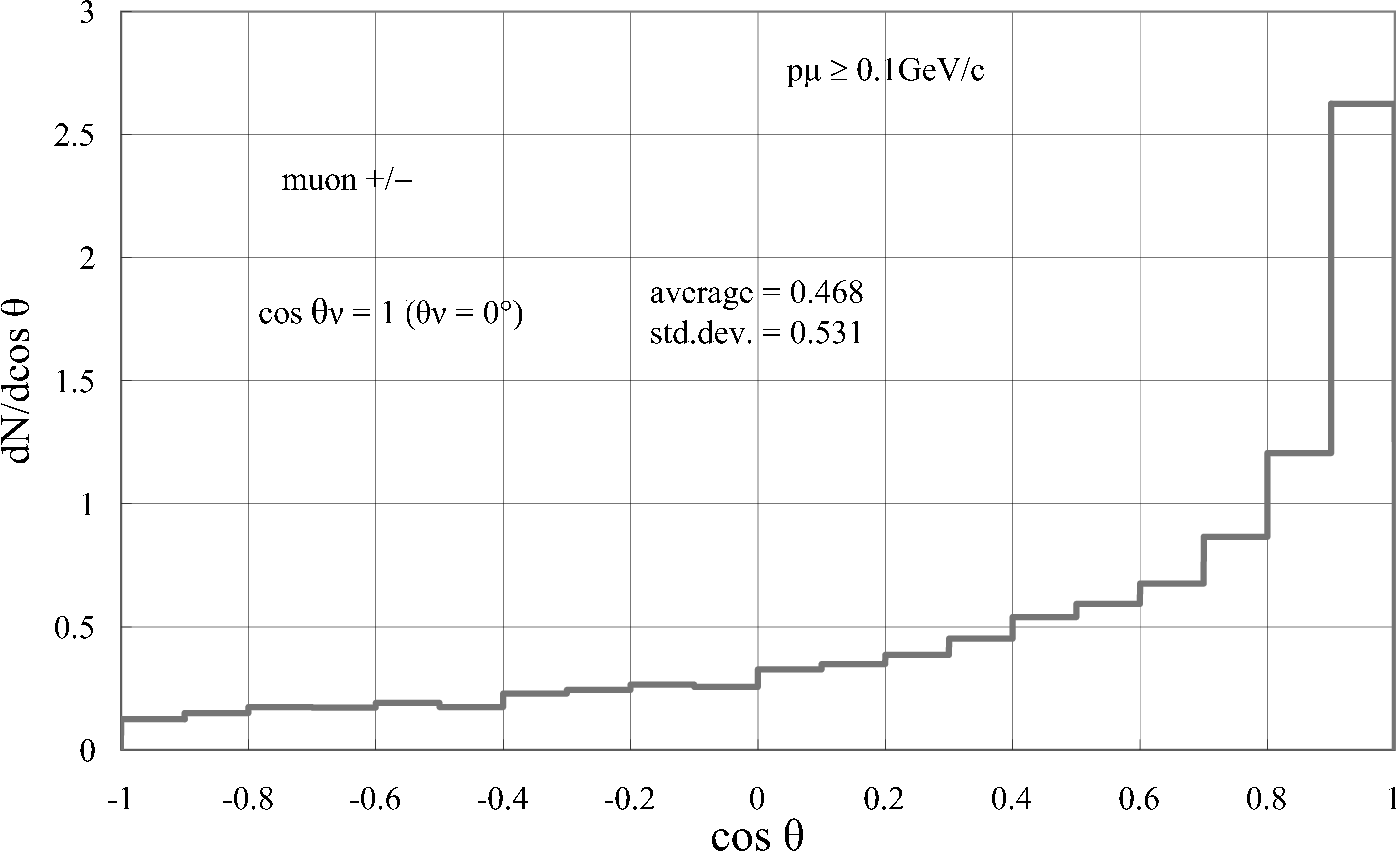}
\vspace{-5mm}
\caption{\label{fig:16} Zenith angle distribution of $\mu^-$ and $\mu^+$ 
for $\nu_{\mu}$ and $\bar{\nu}_{\mu}$ with the vertical direction, 
taking account of the overall neutrino spectrum at Kamioka site.
 The sampling number is 10000.}

\hspace*{10mm}
\includegraphics[scale=0.4,angle=90]{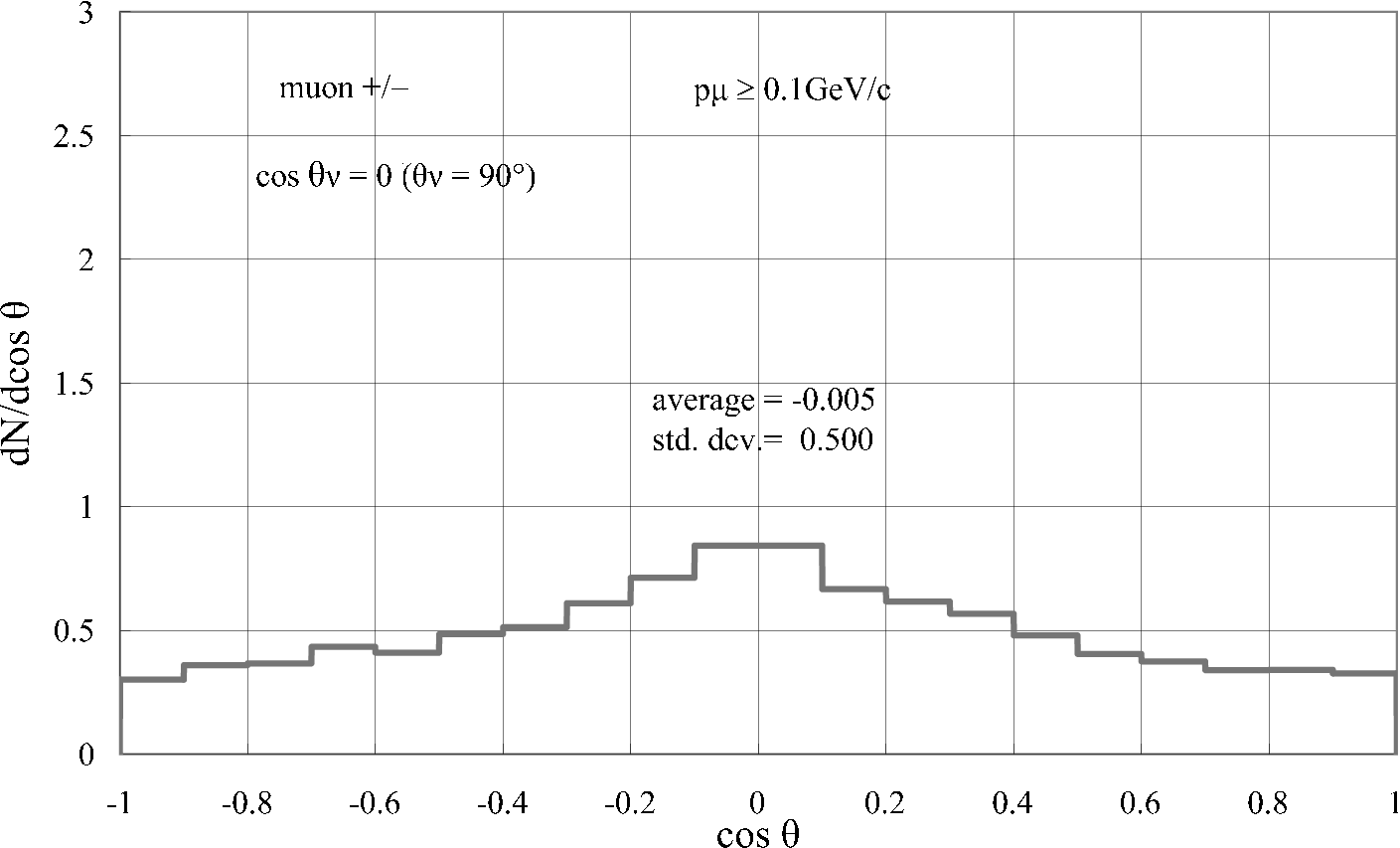}
\vspace{-5mm}
\caption{\label{fig:17} Zenith angle distribution of $\mu^-$ and $\mu^+$ 
for $\nu_{\mu}$ and $\bar{\nu}_{\mu}$ with the horizontal direction, 
taking account of the overall neutrino spectrum at Kamioka site. 
The sampling number is 10000.}

\hspace*{10mm}
\includegraphics[scale=0.4,angle=90]{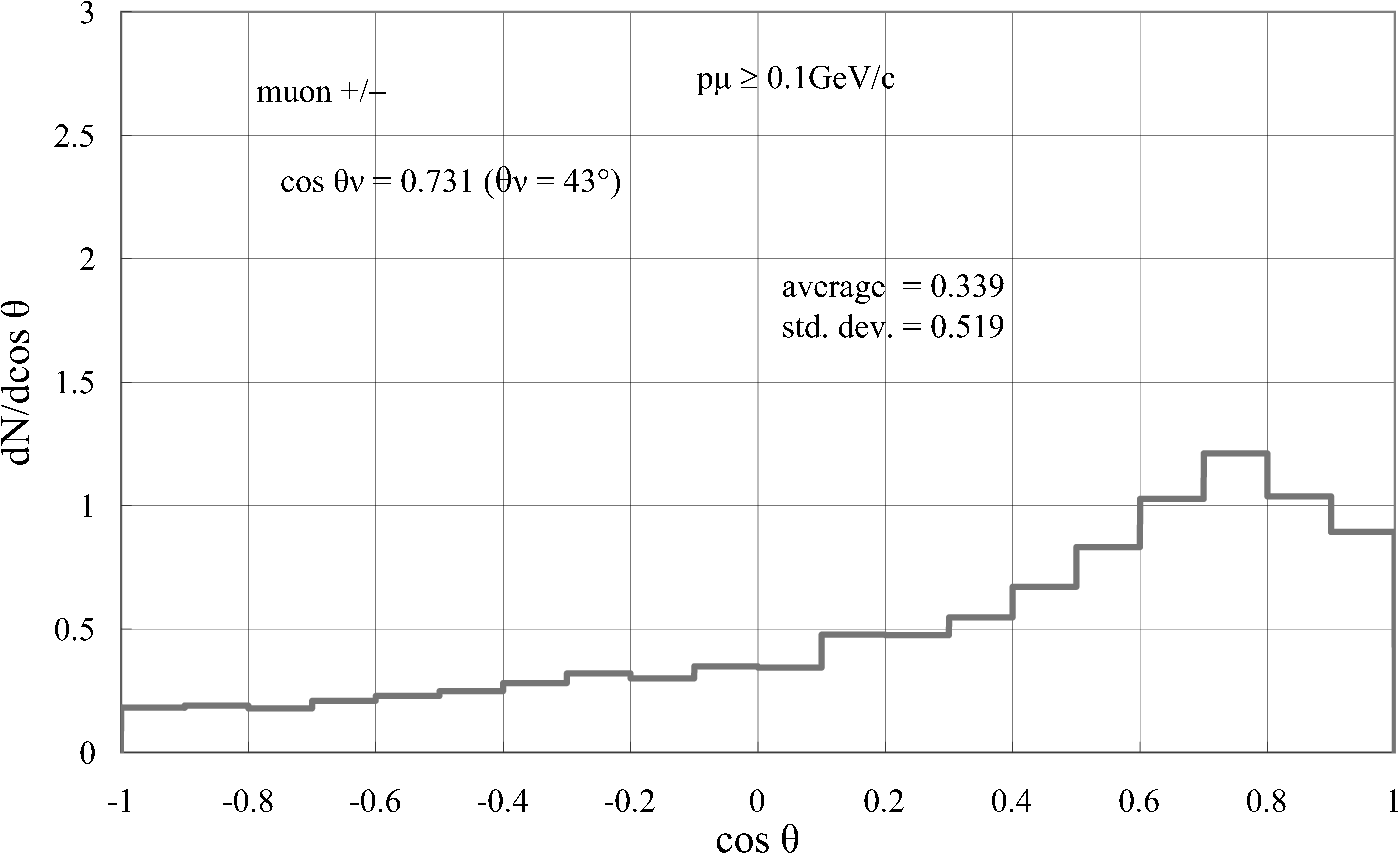}
\vspace{-5mm}
\caption{\label{fig:18} Zenith angle distribution of $\mu^-$ and $\mu^+$ 
for $\nu_{\mu}$ and $\bar{\nu}_{\mu}$ with the diagonal direction, taking 
account of the overall neutrino spectrum at Kamioka site. 
The sampling number is 10000.}
\end{figure}

In Figs. 16 to 18, we give the zenith angle distributions of the sum of 
$\mu^+$ and $\mu^-$ for a given zenith angle of parent neutrinos.
 In Figs. 19 to 21, we give those for $e^+$ 
 and $e^-$ which correspond to  Figs. 16 to 18. There are no major differences
  between
  (anti-)muon and
(anti-)electron. The small differences between them are mainly due to the
different neutrino fluxes between (anti-)muon neutrinos and
(anti-)electron neutrinos.

If the SK assumption is valid, then, the zenith angle distribution of the charged
leptons should be of the delta function type. However, the real distributions 
of the charged leptons in Figs. 16 to 21 are quite far from the delta function 
type distribution.  From Figs. 16 to 21, we can
conclude that the present SK analysis procedures for {\it Fully
Contained Events} and {\it Partially Contained Events} can not essentially
determine the zenith angles of the incident neutrinos.
It should be finally noticed that we must consider not only the contribution of the downward neutrino events which originate from the upward going neutrinos, but also that of the upward neutrino events which originate from the downward going neutrinos.

%

\begin{figure}
\vspace{-37mm}
\hspace*{10mm}
\includegraphics[scale=0.4,angle=90]{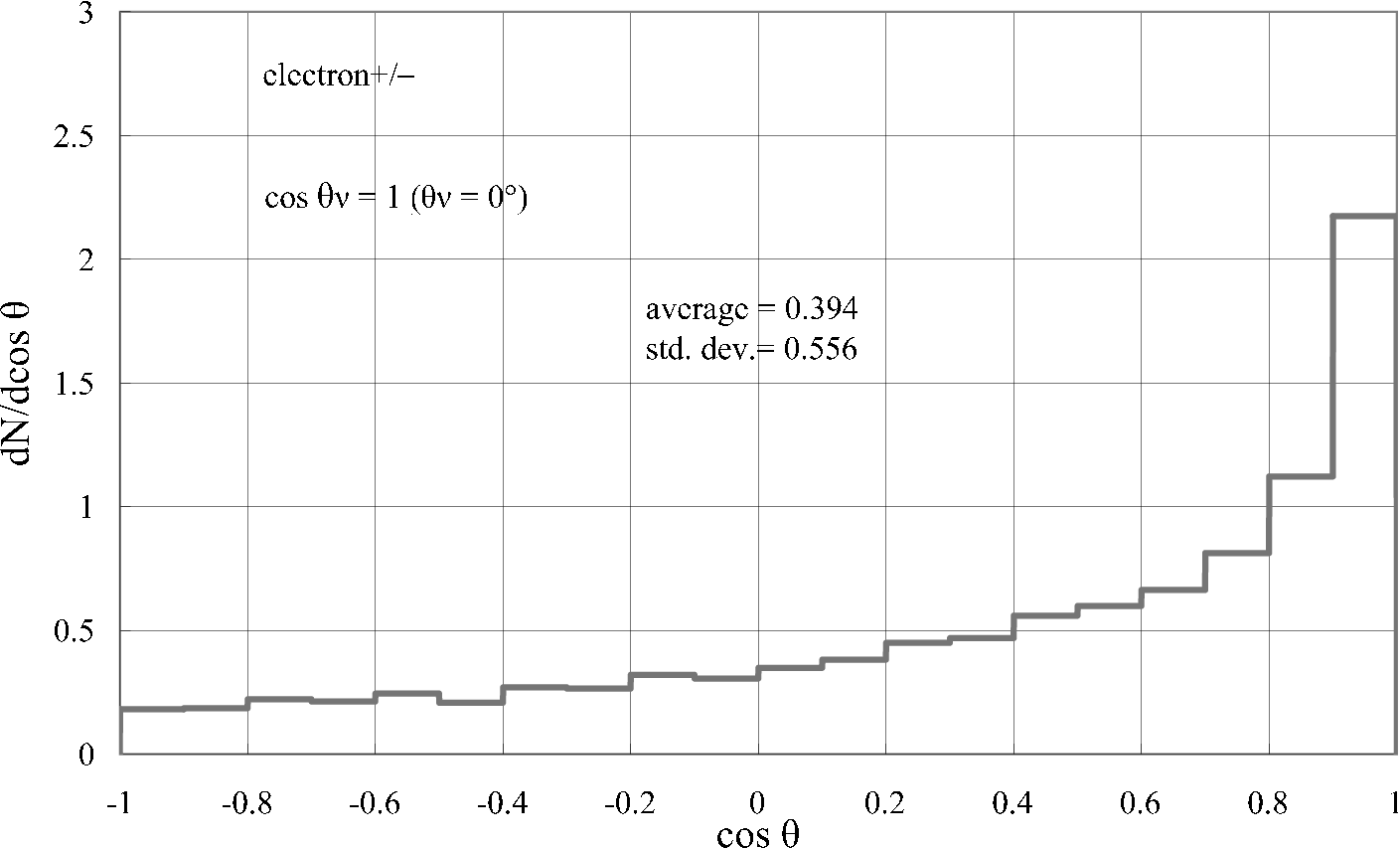}
\vspace{-5mm}
\caption{\label{fig:19} Zenith angle distribution of $e^-$ and $e^+$ for 
$\nu_e$ and $\bar{\nu}_e$ with the vertical direction, taking account of 
the overall neutrino spectrum at Kamioka site. The sampling number is 10000.}

\hspace*{10mm}
\includegraphics[scale=0.4,angle=90]{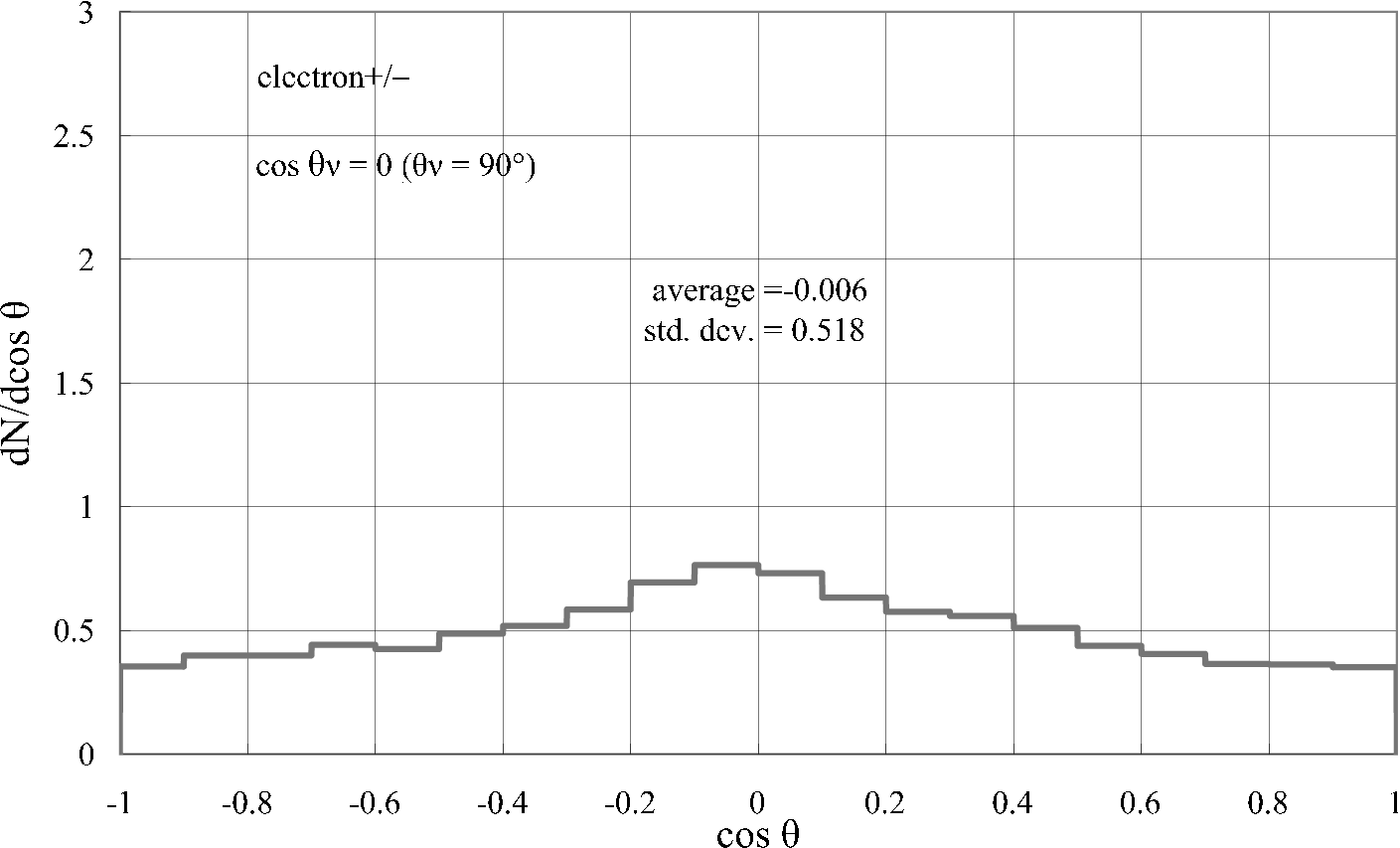}
\vspace{-5mm}
\caption{\label{fig:20} Zenith angle distribution of $e^-$ and $e^+$ 
for $\nu_e$ and $\bar{\nu}_e$ with the horizontal, taking account of 
the overall neutrino spectrum at Kamioka site. The sampling number is 10000.}

\hspace*{10mm}
\includegraphics[scale=0.4,angle=90]{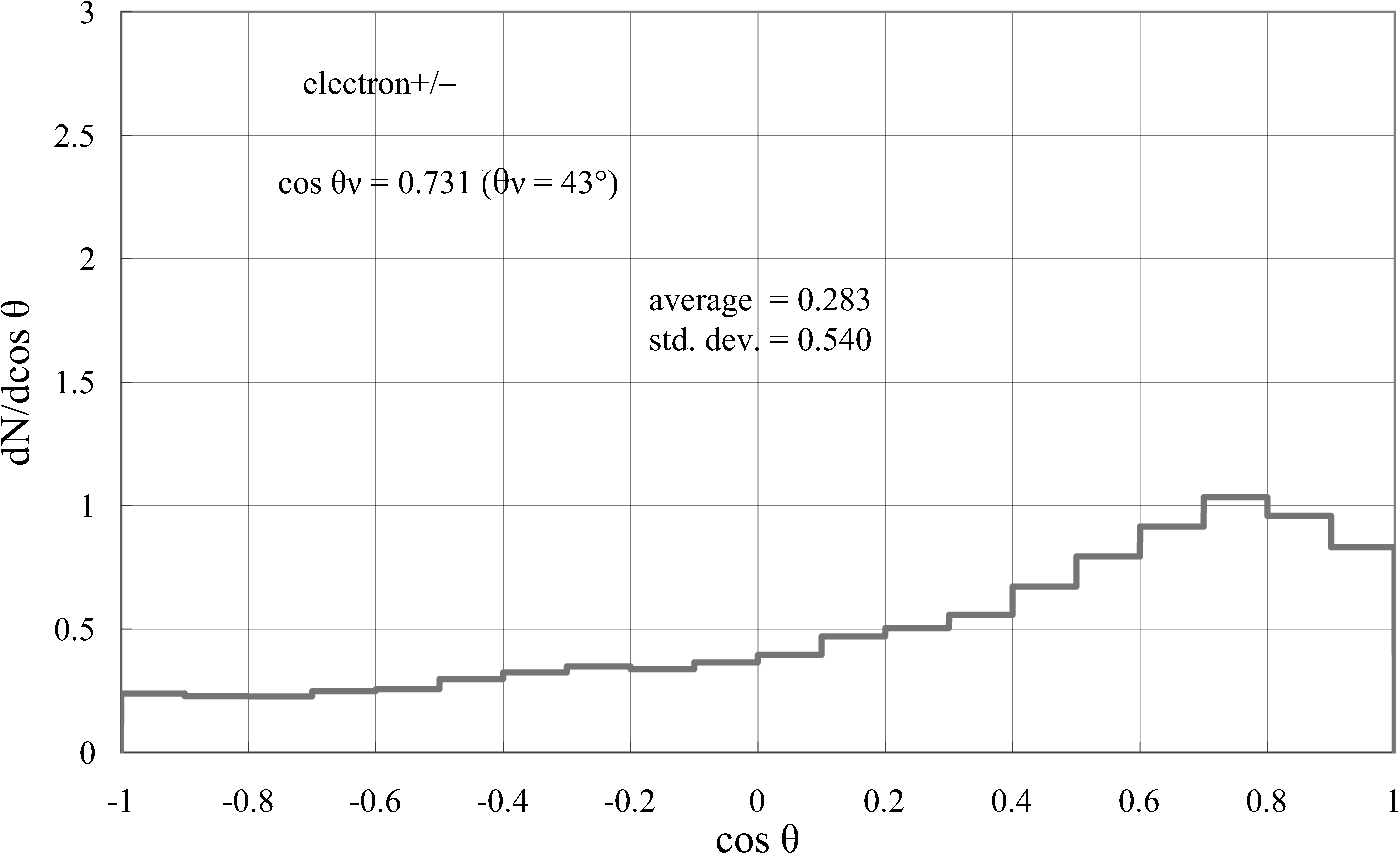}
\vspace{-5mm}
\caption{\label{fig:21} Zenith angle distribution of $e^-$ and 
$e^+$ for $\nu_e$ and $\bar{\nu}_e$ with the diagonal, taking account of 
the overall neutrino spectrum at Kamioka site. The sampling number is 10000.}
\end{figure}

\section{Conclusions}

The standard SK analysis 
for {\it Fully Contained Events} and {\it Partially Contained Events}
 assumes that the direction of
the emitted charged lepton is the same as that of the incident
neutrino. 
However,
Figs. 16 to 22 clearly show that the zenith angle
distributions of the charged leptons have a wide spread for a fixed
zenith angle of the incident (anti-)neutrino, with the leptons even
being scattered backward in some cases, while the corresponding distributions 
obtained from the SK assumption should be of the delta function type, 
the peak of which
  correspond to the zenith angles of the incident neutrinos.
  
  Thus, it is clearly shown that the real zenith 
  angle distributiuon of the charged particle in which 
  their scattering angle is correctly taken into account 
  cannot be approximated in any sense to the delta function 
  type distribution which comes directly from the SK assumption.
   
Consequently, the SK
assumption
 $( n_r  = n )$
does not reflect the real zenith angle for 
directions of the incident neutrinos.




 Namely, our results show clearly that
the zenith angle distribution of the muon-like (single ring) events
and electron-like (single ring) events cannot reflect that of incident
 neutrinos, even if the discrimination
between electron and muon is perfect.
\footnote{We have studied the SK
discrimination between electrons and muons by simulating electron-like
(single ring) events and muon-like (single ring) events and
examining the discrimination between them, based on the concept of
pattern recognition. We conclude that the level of discrimination
claimed by SK is practically impossible to achieve. This study will be
described in a later publication.}

The zenith angle distribution of both {\it Fully Contained
Events} and {\it Partially Contained Events} in our numerical computer
experiment in which
the polar angle (the scattering angle)
 as well as the azimuthal angle of the charged leptons are
correctly taken into account will be published elsewhere.

\end{document}